\documentclass[jkps,preprint,fleqn,showpacs,showkeys]{revtex4}
\usepackage{graphicx}
\usepackage{amssymb}
\usepackage{amsmath}
\usepackage{bm}
\begin{document}
\setcounter{page}{0}
\title[]{ASTROPHYSICAL AND LABORATORY PLASMAS: HF PROPERTIES UNDER EXTREME CONDITIONS}
\author{Vladimir A. \surname{Sre\'ckovi\'c}}
\email{vlada@ipb.ac.rs}
\author{Anatolij.A. \surname{Mihajlov}}
\author{Nenad. M. \surname{Sakan}}
\author{Ljubinko.M. \surname{Ignjatovi\'c}}
%\thanks{Fax: +381 (0)11 31 62 190}
\affiliation{Institute of physics, University of Belgrade, P.O. Box 57 11001, Belgrade, Serbia}
\author{Darko \surname{Jevremovi\'c}}
\author{Veljko \surname{Vuj\v ci\'c}}
\author{Milan S. \surname{Dimitrijevi\'c}}
\affiliation{Astronomical Observatory, Volgina 7 11060, Belgrade 74, Serbia}

\date[]{Received 6 August 2007}

\begin{abstract}
The values of electrical conductivity of plasma of stars with a magnetic field or moving in the magnetic field of the other component in a binary system could be of significant interest, since they are useful for the study of thermal evolution of such objects, cooling, nuclear burning of accreted matter, and the investigation of their magnetic fields.  So, on the basis of numerically calculated values for the dense plasma conductivity in an external HF electric field, we determine the HF characteristics of astrophysical plasmas under extreme conditions. The examined range of frequencies covers the IR, visible and near UV regions and consider electronic number density and temperature are in the ranges of $10^{21} \textrm{cm}^{-3}  \le Ne$ and 20 000 $\le T$, respectively. The method developed here represents a powerful tool for research into white dwarfs with different atmospheric compositions (DA, DC etc.), and for investigation of some other stars (M-type red dwarfs, Sun etc.).
\end{abstract}



\maketitle

\section{INTRODUCTION}
 Exploring and improving the new
calculation possibilities, simulation techniques and the extension of
numerous models in connection with the dynamic properties of nonideal plasma is
in the focus of investigators nowadays \cite{bai95,maz07, fre12,col13,for10}.

In this paper it is considered a highly ionized plasma in
a homogenous and monochromatic external electric field
$\vec{E}(t) = \vec{E_0} \exp \left\{ -i \omega t \right\}$.
According to \cite{ada04}, the dynamic electric conductivity of
a strongly coupled plasma $\sigma (\omega )=\sigma _{Re}(\omega )+i\sigma _{Im}(\omega )$
is presented by the expressions

\begin{equation}
  \sigma (\omega ) = \frac{4e^{2}}{3m} \int_{0}^{\infty} \tau(E)
  \left[- \frac{dw(E)}{dE}\right] \rho (E) E dE,  \label{eq::sig_rpa}
\end{equation}

\begin{equation}
  \sigma _{Re}(\omega )=\frac{4e^{2}}{3m}\int_{0}^{\infty }\frac{\tau (E)}{
    1+(\omega \tau (E))^{2}}\left[ -\frac{dw(E)}{dE}\right] \rho
  (E)EdE \label{eq::sig_rpa_Re}
\end{equation}

\begin{equation}
  \label{eq::sig_rpa_Im}
  \sigma_{Im}(\omega) = \frac{4e^{2}}{3m} \int_{0}^{\infty}\frac{
    \omega \tau^2(E)}{1 + (\omega \tau(E))^2}
  \left[-\frac{dw(E)}{dE}\right] \rho(E) E dE
\end{equation}

{\noindent where $\rho (E)$ is the density of a electron states in the
  energy space and $w(E)$ is the Fermi-Dirac distribution function,
  $\tau(E)$ is the relaxation time}

\begin{equation}
  \tau(E) = \frac{\tau (E)}{1 - i \omega \tau (E)},  \label{eq::tau_rpa}
\end{equation}

{\noindent $\tau (E)$ being the 'static' relaxation time. The method
  of determination of $\tau (E)$ is described in the previous papers
\cite{ada04,dju91,sre10b,sre10c,sre10d,tka06} in detail.}

Other HF plasma characteristics can be expressed in terms of the
quantities $\sigma _{Re}(\omega )$ and $\sigma_{Im}(\omega)$.

Thus the plasma dielectric permeability is

\begin{equation}
  \label{eq::Epsilon_Re_Im}
  \varepsilon (\omega) = 1 + i \frac{4 \pi}{\omega} \sigma (\omega)
  = \varepsilon _{Re} (\omega) + i \varepsilon _{Im} (\omega),
\end{equation}

{\noindent where $\varepsilon _{Re} (\omega)$ and $\varepsilon _{Im}
  (\omega)$ are given as}

\begin{equation}
  \label{eq::Epsilon_Re}
  \varepsilon _{Re} (\omega) = 1 - \frac{4 \pi}{\omega} \sigma_{Im} (\omega) , \quad
  \varepsilon _{Im} (\omega) =  \frac{4 \pi}{\omega} \sigma_{Re} (\omega).
\end{equation}

The coefficients of refraction, $n ( \omega )$, and reflection, $R (
\omega )$, are determined as

\begin{equation}
  \label{eq::Refract}
  n ( \omega ) = \sqrt {\varepsilon (\omega)} = n_{Re} ( \omega ) + i n_{Im} ( \omega ),
\end{equation}

\begin{equation}
  \label{eq::Reflect}
  R(\omega) = \left| \frac{n ( \omega ) - 1}{n ( \omega ) + 1} \right| ^2
\end{equation}

{\noindent where, bearing in mind that}

\begin{equation}
  \label{eq::Moduo_epsilon}
  |\varepsilon (\omega)| = \sqrt{\varepsilon _{Re} ^2 (\omega) + \varepsilon _{Im} ^2 (\omega)} ,
\end{equation}

{\noindent the real and imaginary pert of refractivity, $n_{Re} (
  \omega )$ and $n_{Im} ( \omega )$, are given by}

\begin{equation}
  \label{eq::Refract_Re}
  n_{Re} ( \omega ) = \sqrt{\frac{1}{2} (|\varepsilon (\omega)| + \varepsilon _{Re} (\omega))}, \quad
  n_{Im} ( \omega ) = \sqrt{\frac{1}{2} (|\varepsilon (\omega)| - \varepsilon _{Re} (\omega))}.
\end{equation}

{\noindent From here the equation for the plasma reflectivity could be
  expressed as}

\begin{equation}
  \label{eq::Reflect_detail}
  R(\omega) = \left\{
    \frac{1 + |\varepsilon (\omega)| - \sqrt{2} \sqrt{|\varepsilon (\omega)| + \varepsilon _{Re} (\omega)}}
      {1 + |\varepsilon (\omega)| + \sqrt{2} \sqrt{|\varepsilon (\omega)| + \varepsilon _{Re} (\omega)}}
  \right\} ^{1/2}
\end{equation}

The other parameter of interest is the penetration depth
of electromagnetic radiation into plasma, $\Delta(\omega)$.
This quantity is just the skin-layer width
determined as the inverse imaginary part of the electromagnetic field wave number

\begin{equation}
  \label{eq::Penetracija}
  \Delta (\omega) = \frac{c}{\omega} \frac{1}{n_{Im} (\omega)}.
\end{equation}

{\noindent where $c$ is the speed of light.}

\section{RESULTS AND DISCUSSION}

We here continue our previous investigations of plasma static electrical conductivity
which are of interest for DB white dwarf atmospheres (see e.g. \cite{sre10a, ada06}).

So, in accordance with the aim of this work, we calculated HF plasma characteristics for vide plasma conditions
in order to apply our results on the atmospheres of different stellar types.

\begin{figure}
\centering
\includegraphics[height=0.34\textwidth]{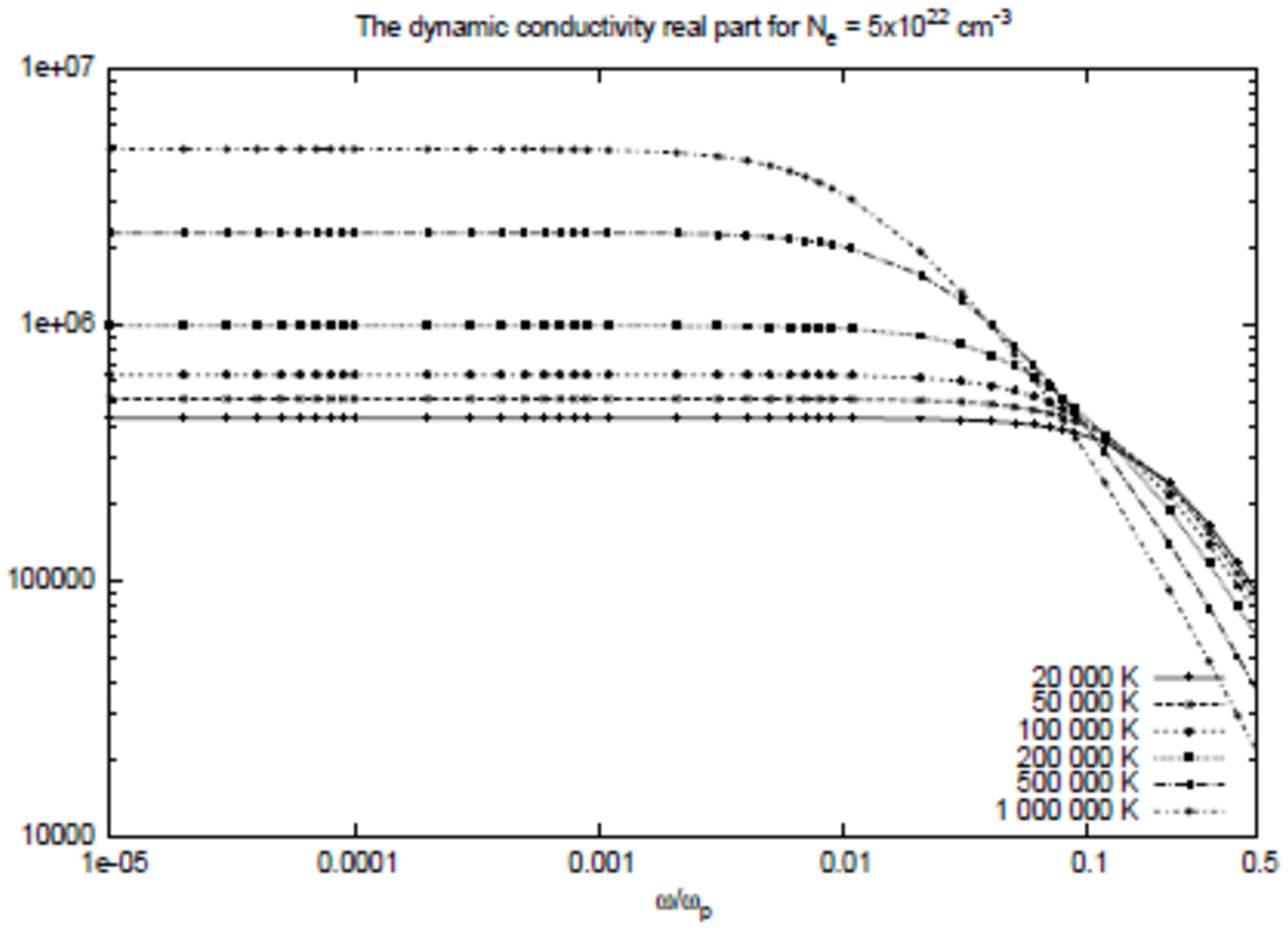}
\includegraphics[height=0.34\textwidth]{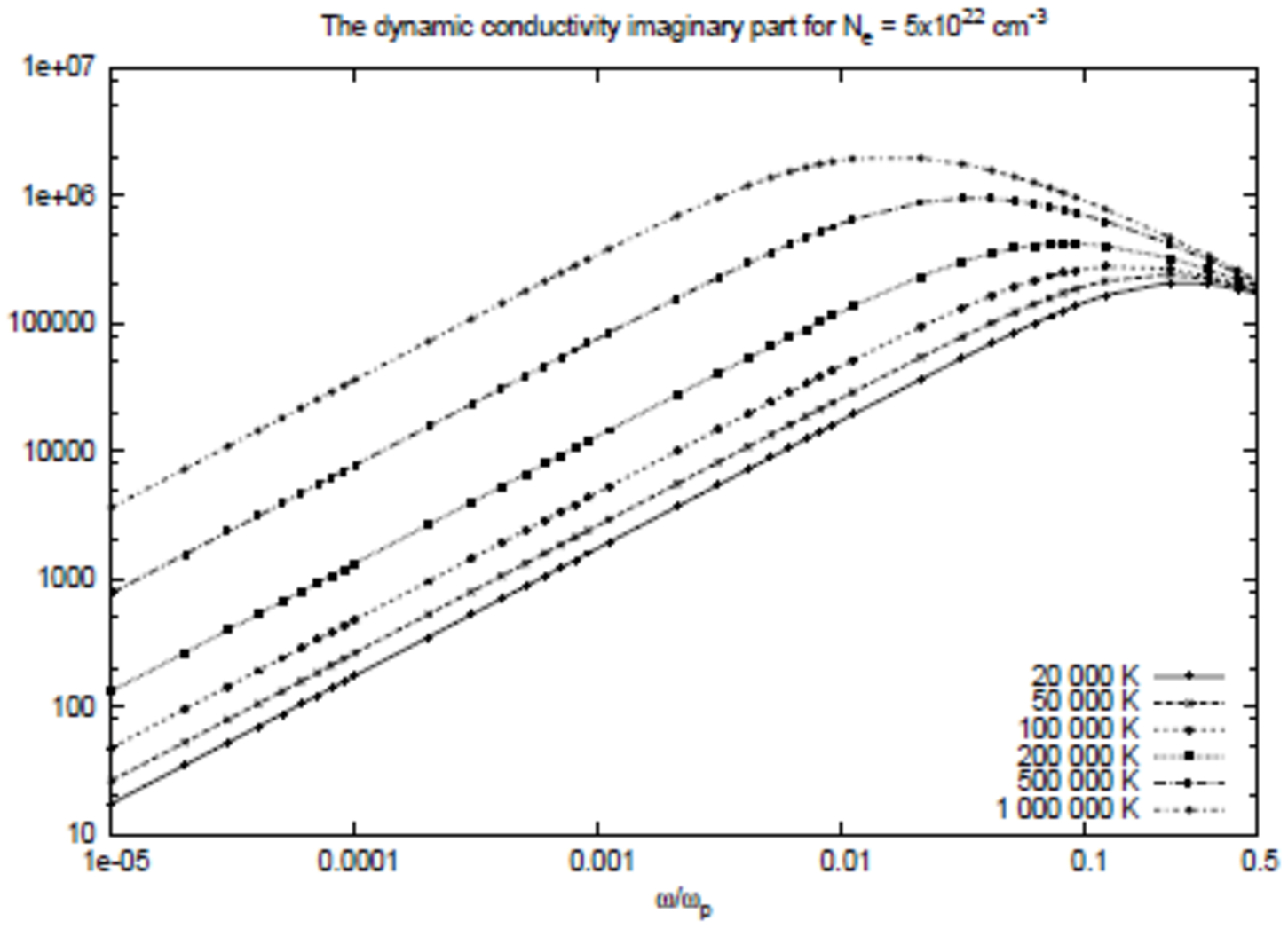}
\caption{The dynamic conductivity real $\sigma _{Re}(\omega)$ and imaginary part $\sigma _{Im}(\omega)$
for $Ne = 5\cdot10^{22}$ cm$^{-3}$ and $20000 \textrm{K} <T<100000 \textrm{K}$.}
\label{fig:1}
\end{figure}

\begin{figure}
\centering
\includegraphics[height=0.34\textwidth]{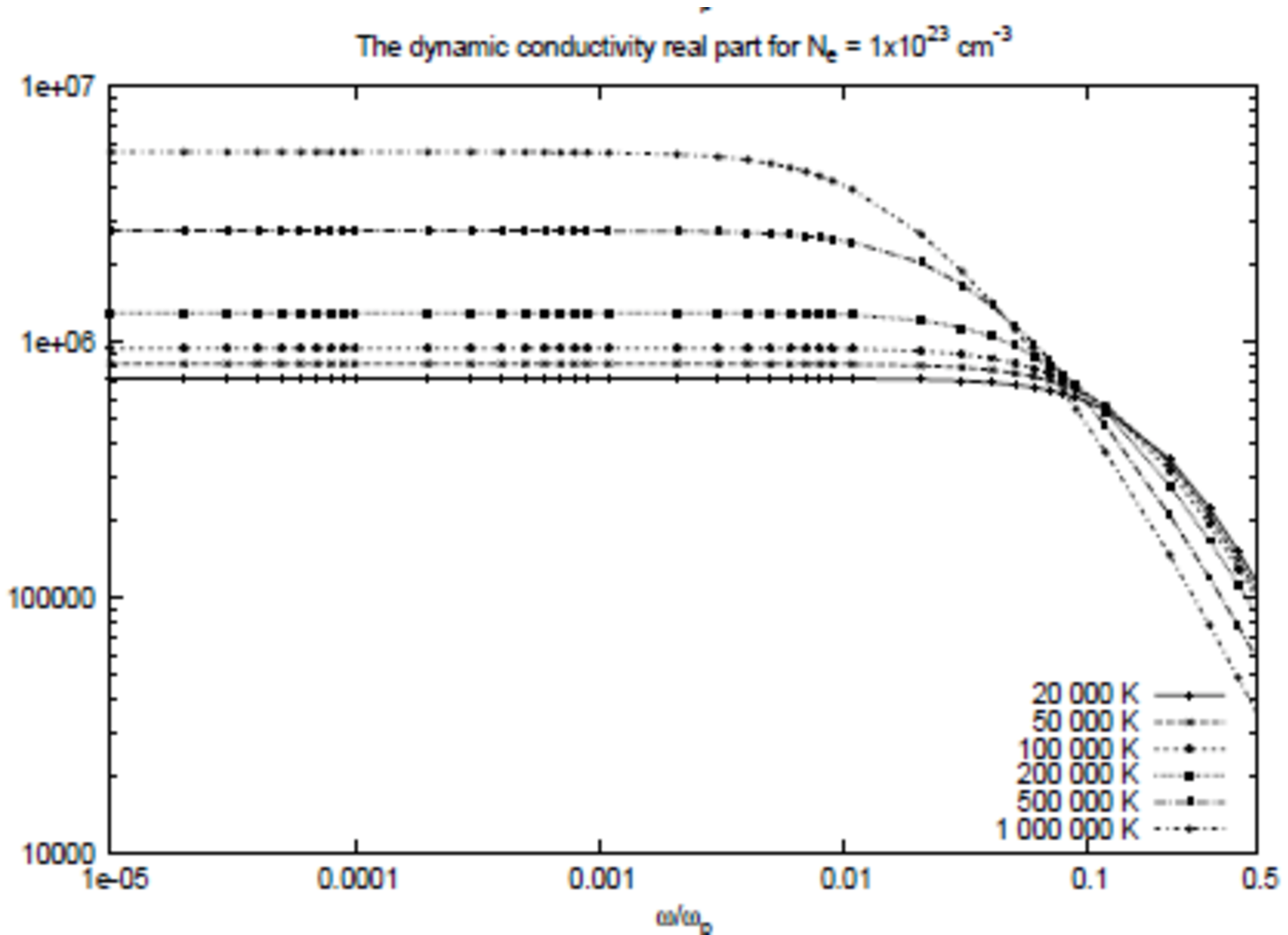}
\includegraphics[height=0.34\textwidth]{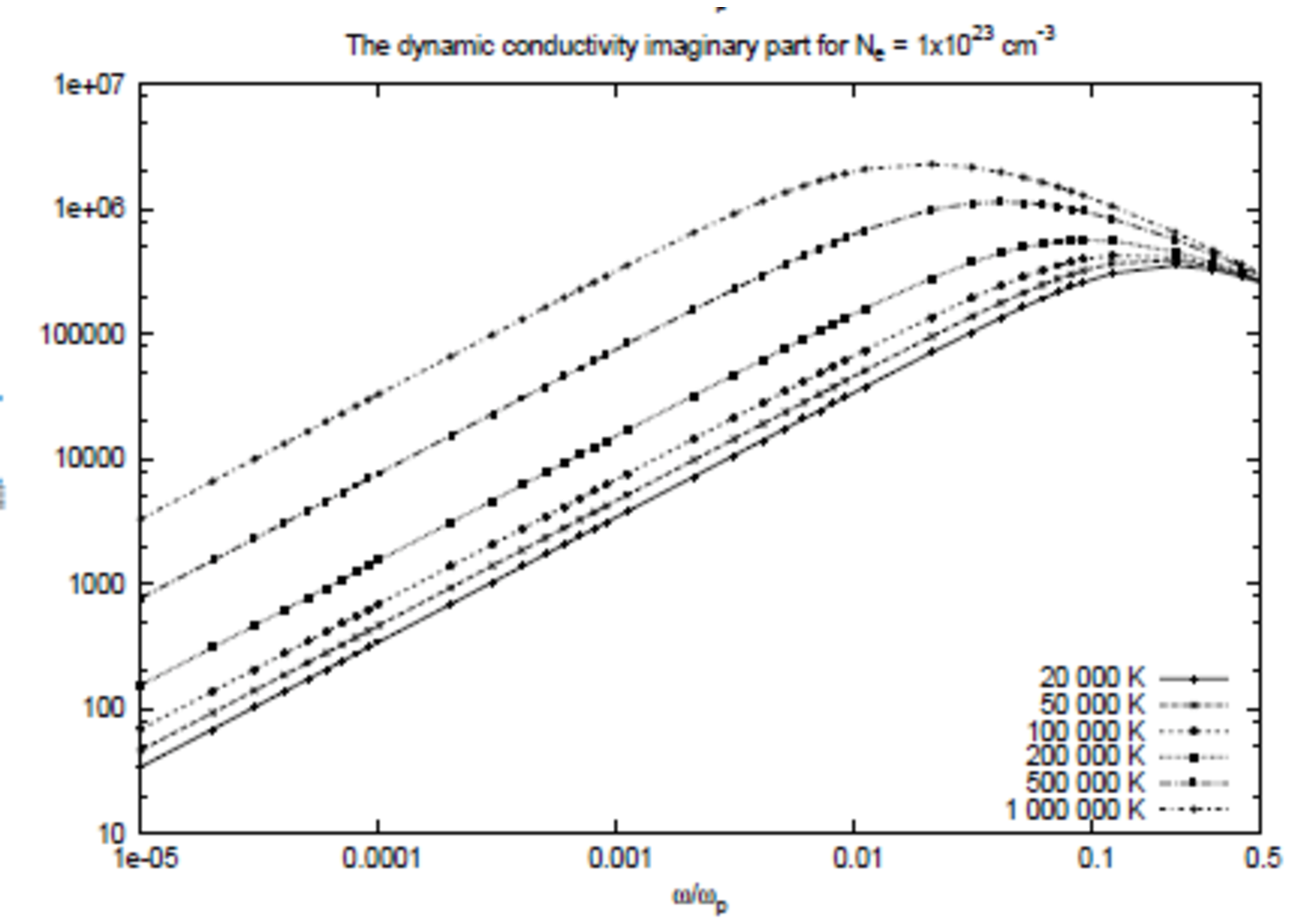}
\caption{The dynamic conductivity real $\sigma _{Re}(\omega)$ and imaginary part $\sigma _{Im}(\omega)$
for $Ne = 1\cdot10^{23}$ cm$^{-3}$ and $20000 \textrm{K} <T<100000 \textrm{K}$.}
\label{fig:2}
%%%%%%%%%%%%%%%%%%%%%%%%%%%%
\end{figure}

\begin{figure}
\centering
\includegraphics[height=0.34\textwidth]{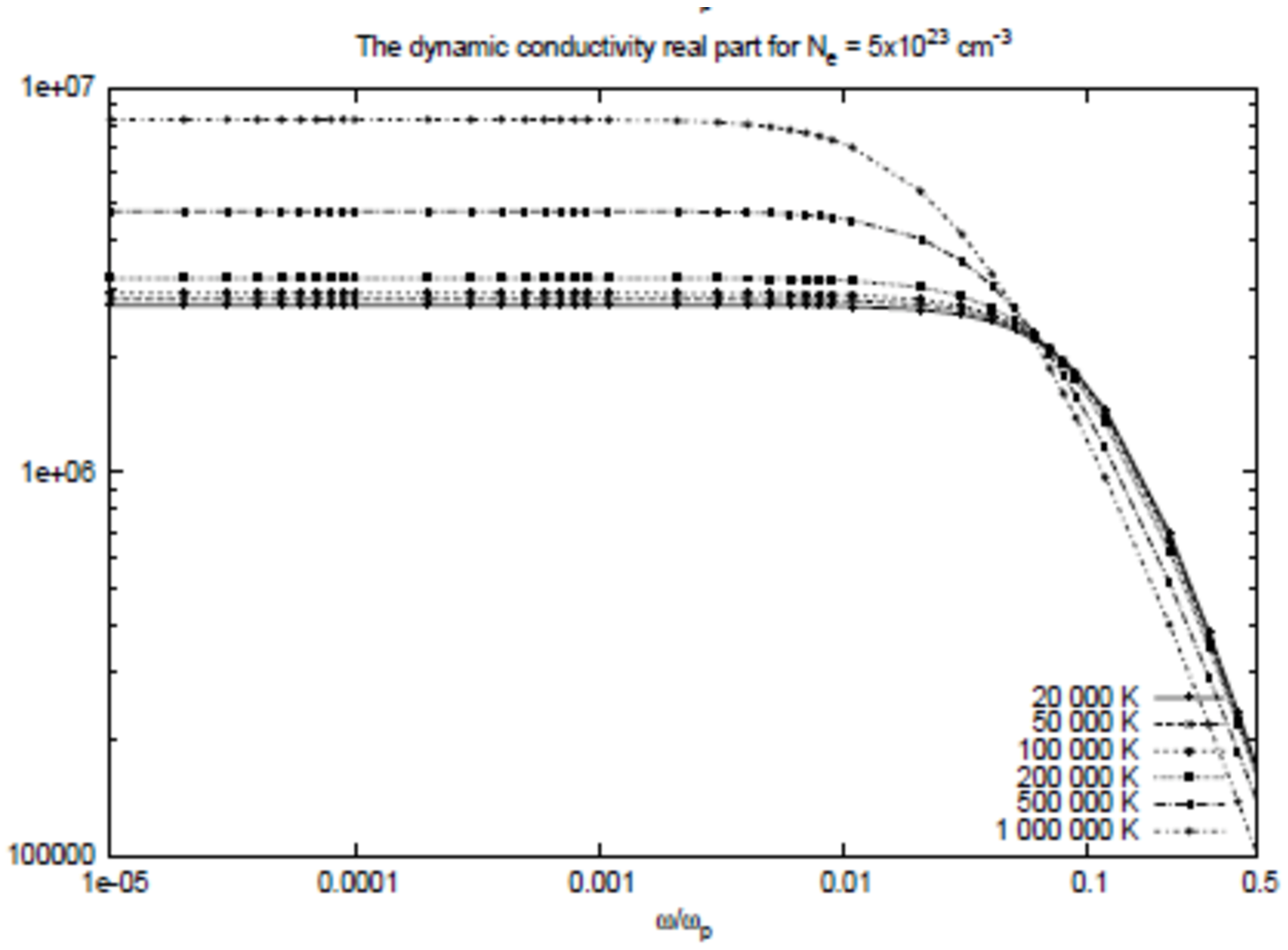}
\includegraphics[height=0.34\textwidth]{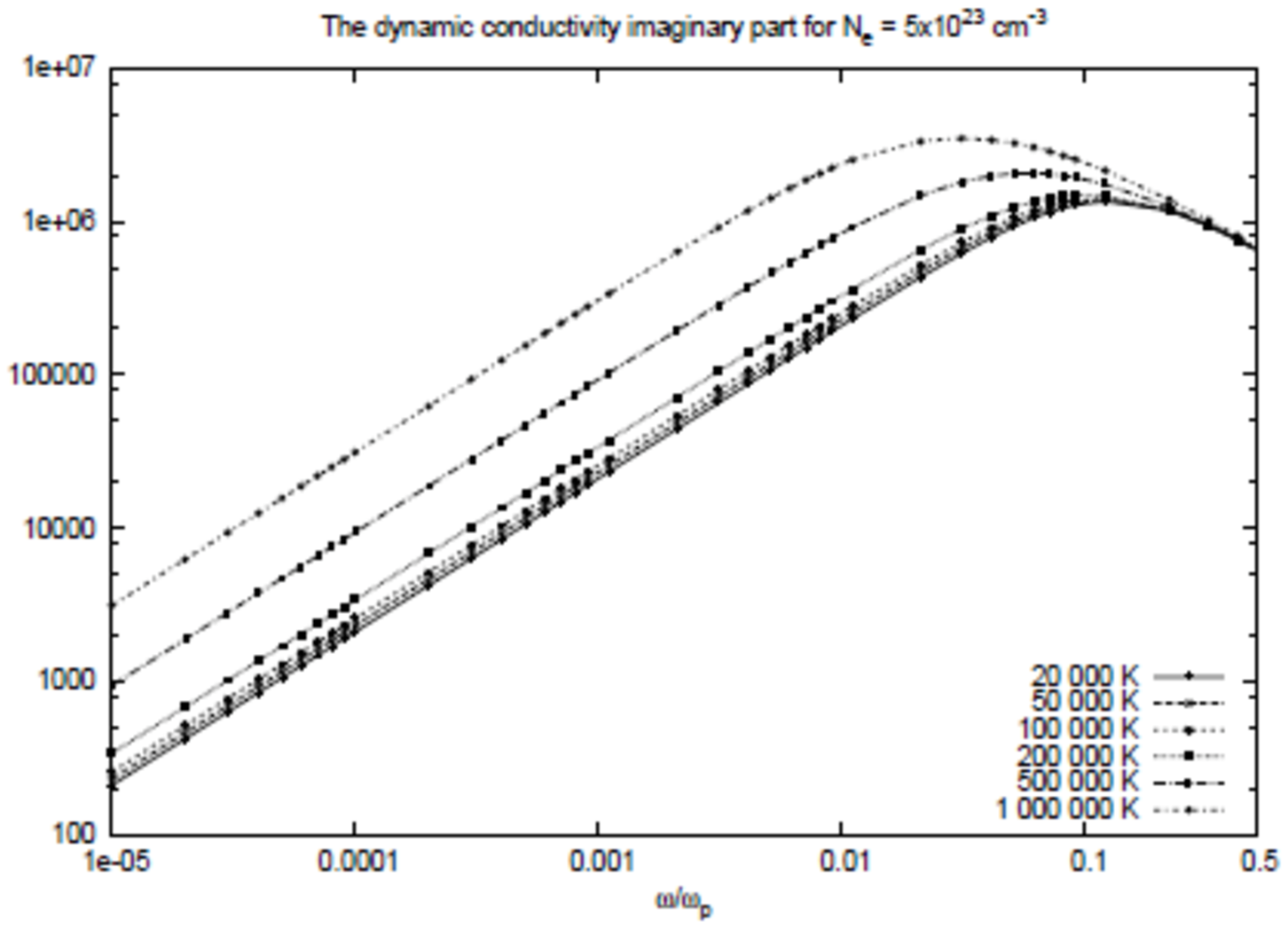}
\caption{The dynamic conductivity real $\sigma _{Re}(\omega)$ and imaginary part $\sigma _{Im}(\omega)$
for $Ne = 5\cdot10^{23}$ cm$^{-3}$ and $20000 \textrm{K} <T<100000 \textrm{K}$.}
\label{fig:3}
\end{figure}

Figures \ref{fig:1}-\ref{fig:3} illustrates the behavior of the HF conductivity for various plasma conditions
which gives possibility to calculate other transport properties.
The figures \ref{fig:1}-\ref{fig:3}, demonstrate the regular behavior of $\sigma
_{Re}(\omega )$, i.e. the convergence to the corresponding values of
$\sigma _{0}(n_{e},T)$ when $\omega \rightarrow 0$, and the
existence of the interval of variation of $\omega $ where $\sigma
_{Re}(\omega)$ is practically constant. We observe the tendency of
this interval to decrease when temperature $T$ increases. Similarly,
the figures \ref{fig:1}-\ref{fig:3}, demonstrate a regular behavior of
$\sigma_{Im}(\omega)$, i.e. the convergence to zero when $\omega
\rightarrow 0$, and the presence of a maximum in the interval
$0<\omega <0.5\omega _{p}$. 
%\vfil\newpage

% Figure (in PS or EPS format)
%%%%%%%%%%%%%%%%%%%%%%%%%%%%
%%%%%%%%%%%%%%%%%%%%%%%%%%%%%%

Our plane is to present the results obtained during this investigation
in database which can be accessed directly through
http://servo.aob.rs as a web service similarly to the existing MOL-D and E-MOL databases http://servo.aob.rs/mold,
http://servo.aob.rs/emol/ (see e.g. \cite{mar15,vuj15}).

The method developed in this paper represents a powerful
tool for research white dwarfs with different atmospheric
compositions (DA, DC etc.), and some other stars (M-type red dwarfs, Sun etc.).
Finally, the presented method provides a basis for the development of methods
to describe other transport characteristics which are important for the
study of all mentioned astrophysical objects, such as the electronic
thermo-conductivity in the stellar atmosphere layers with
large electron density, and electrical conductivity in the presence of
strong magnetic fields or plasma reflectivity \cite{dav12} in high energy and high density plasma.

\begin{acknowledgments}
The authors are thankful to the MESTD of RS for support
of this work within projects 176002 and III44002.
\end{acknowledgments}

%\textbf{SUPPORTING INFORMATION}:\\
%Additional Supporting Information (Tables) may be found in the on-line version of this article.

\end{document}